\documentclass[letterpaper,journal]{IEEEtran}
\usepackage[T1]{fontenc}

\usepackage{amsmath,amsfonts,amssymb,bm}
\usepackage{graphicx}
\usepackage{cite}
\usepackage{url}
\usepackage{stfloats}
\usepackage{placeins}
\usepackage{booktabs}
\usepackage{array}
\usepackage[caption=false,font=footnotesize]{subfig}
\usepackage{enumitem}
\usepackage{comment}
\usepackage{xcolor}
\usepackage{multirow}
\usepackage{tabularx}

\setkeys{Gin}{trim={5mm 10mm 5mm 8mm},clip}

\graphicspath{{FIG/}{fig/}}
\newenvironment{revision}{\color{black}}{}
\newcommand{\rev}[1]{{\color{black}#1}}

\newcommand{\RR}{\mathbb{R}}
\newcommand{\dd}{\mathrm{d}}

\newcommand{\norm}[1]{\left\lVert #1 \right\rVert}

\newcommand{\Frenet}{\mathbf{F}}
\newcommand{\OmegaF}{{\mathbf \Omega}_{\scriptscriptstyle \rm F}}

\newcommand{\tnb}{\scriptscriptstyle \rm TNB}

\newcommand{\iT}{i_{\scriptscriptstyle \rm T}}
\newcommand{\iN}{i_{\scriptscriptstyle \rm N}}
\newcommand{\iB}{i_{\scriptscriptstyle \rm B}}
\newcommand{\T}{{\scriptscriptstyle \rm T}}
\newcommand{\N}{{\scriptscriptstyle \rm N}}
\newcommand{\B}{{\scriptscriptstyle \rm B}}
\newcommand{\PI}{{\scriptscriptstyle \rm PI}}

\newcommand{\plotbox}[2]{%
\includegraphics[width=#2\linewidth]{#1}%
}

\begin{document}

\title{PLL-Less Synchronization of Grid-Following Converters Using a Voltage-Driven Frenet Frame}

\author{Jinhui~Chen,~\IEEEmembership{Student Member,~IEEE,}
        Huadong~Sun,~\IEEEmembership{Senior Member,~IEEE,}
        Shiyun~Xu,~\IEEEmembership{Senior Member,~IEEE,}
        Bing~Zhao,~\IEEEmembership{Senior Member,~IEEE,}
        Peng~Han,
        and~Federico~Milano,~\IEEEmembership{Fellow,~IEEE}%

\thanks{This work was supported by the Smart Grid National Science and Technology Major Project, China, under Grant 2025ZD0806900. \textit{(Corresponding author: Huadong Sun.)}}
\thanks{J.~Chen, H.~Sun, S.~Xu, B.~Zhao, and P.~Han are with the China National Key Laboratory of Power Grid Security, China Electric Power Research Institute, Beijing, China. F.~Milano is with University College Dublin, Dublin, Ireland.}
}

\markboth{IEEE Transactions on Energy Conversion}%
{Chen \MakeLowercase{\textit{et al.}}: PLL-Less Frenet-Frame Synchronization and Current Control}

\maketitle
\bstctlcite{BSTcontrol}


\begin{abstract}
In grid-following converters, the synchronization path can transmit
voltage disturbances to the current loop under weak or unbalanced grid
conditions. This paper develops a Frenet-frame controller that derives
its axes from the integral curve of the measured voltage at the point of
common coupling. The tangential current governs active power, while the
normal and binormal currents govern the transverse components of
non-active power. Feedforward terms derived from the output filter model
compensate curvature and torsion coupling, yielding identical nominal
current dynamics across the axes. The controller operates without a
phase-locked loop or sequence extraction. A reduced transverse model
accounts for the finite bandwidth of geometric rate reconstruction and
predicts a reduction in low-frequency coupling from voltage disturbances
to current. Controller hardware-in-the-loop tests using real-time
simulation of a three-phase four-wire converter show the shortest
settling time after an active power reference step and the smallest peak
power deviations during balanced voltage disturbances among the tested
controllers. Under single-phase voltage unbalance, the method reduces
peak binormal current and maximum phase-current total harmonic distortion
relative to both benchmarks. It also settles to the commanded power in a
weak-grid connection test where the phase-locked loop benchmark exhibits
sustained oscillations.
\end{abstract}


\section{Introduction}

\IEEEPARstart{G}{rid}-following (GFL) voltage-source converters
are widely used as interfaces for renewable generation, battery energy
storage, and other inverter-based resources because of their mature
implementation and clear control structure
\cite{blaabjerg2006_overview,golestan2017_pllreview,
cespedes2014_impedance}.
Most implementations rely on phase-locked loop (PLL) synchronization and
$dq$-frame current regulation
\cite{jaalam2016_review,ali2018_pllreview,rodriguez2007_ddsrf}.
Under balanced sinusoidal conditions, this structure transforms
three-phase ac variables into approximately constant quantities in the
synchronous frame and thereby simplifies current and power regulation.
However, PLL synchronization and the associated voltage processing become
increasingly challenging under weak grids, voltage unbalance, waveform
distortion, and rapid transients
\cite{entsoe2025_iberia,derviskadic2021_beyondphasors,
lara2024_tdsim}.

\begin{revision}
Synchronization and estimation structures based on the second-order
generalized integrator (SOGI) have been developed for nonideal voltage
conditions, including SOGI-based PLLs
\cite{ciobotaru2006_sogi} and dual second-order generalized-integrator
frequency-locked loops (DSOGI-FLLs)
\cite{rodriguez2006_dsogifll,rodriguez2012_dsogifll}.
These methods improve synchronization-variable and voltage-component
estimation under unbalance and distortion, but retain a dedicated
voltage-processing stage between the measured grid voltage and current
regulation.
\end{revision}

\begin{revision}
A different research direction removes the explicit PLL from the GFL
control structure.
Voltage-modulated direct power control regulates instantaneous
active and reactive powers directly in stationary coordinates
\cite{gui2019_vmdpc,gui2019_gvmdpc_distorted,
gong2023_vmdpc_decoupling}.
Linear parameter-varying power-synchronized control (LPV-PSC)
establishes synchronization through active-power dynamics
\cite{mansour2022_lpv_psc,mansour2024_transient}, while other PLL-less
approaches employ oscillator-based synchronization, predictive control,
or reactive-power synchronization
\cite{awal2021_uvoc,dong2023_pllless_deadbeat,
tang2026_reactive_sync}.
However, removing the PLL alone does not eliminate the dynamic interaction
between voltage processing, synchronization, and current regulation
\cite{hoseinizadeh2022_pllless_vcc}.
A further issue is therefore how the current-control coordinates and the
associated power channels are constructed from the measured three-phase
voltage.
\end{revision}

\begin{revision}
Conventional GFL controllers generally employ stationary coordinates or
prescribed planar rotating coordinates.
Such representations are well suited to conventional active- and
reactive-power regulation.
Under general nonideal three-phase conditions, however, the integral
curve generated by the three-phase voltage is not guaranteed to remain
in a fixed plane.
In a prescribed planar representation, additional components of the
multidimensional instantaneous non-active-power vector generally remain
functions of the same in-plane current variables used for conventional
active- and reactive-power regulation.
They are therefore not explicitly separated into distinct orthogonal
current-control channels.
This motivates a control representation generated directly from the
instantaneous three-phase voltage trajectory that explicitly separates
the multidimensional power components within the current-control
coordinates.
\end{revision}

Alternative coordinate and frequency representations have been developed
for nonideal three-phase conditions.
The Tan--Sun transformation
\cite{tan2017_tansun} maps unbalanced ac quantities into dc quantities
without second-harmonic components, whereas the generalized g-$dq0$
framework
\cite{tayyebi2024_gdq0} provides time-invariant coordinates for
prescribed phase- and magnitude-unbalanced waveforms.
The complex-frequency framework
\cite{milano2022_complexfreq} describes normalized voltage-magnitude and
angular-frequency dynamics and has been further applied to converter
synchronization and closed-loop estimation
\cite{he2025_cfsync,quan2025_cfll}.
These approaches provide useful representations of nonideal voltage
dynamics, but they do not establish the voltage-trajectory-driven three-dimensional coordinate representation and multidimensional
power-channel structure considered in this paper.

The geometric-frequency framework provides a trajectory-based description
of three-phase electrical quantities.
In \cite{milano2022_geomfreq}, frequency was interpreted through the local
kinematics of three-phase trajectories in $\RR^3$.
The Frenet frame was subsequently shown to generalize the Park transform
by defining tangent, normal, and binormal coordinates directly from the
local trajectory without an externally specified synchronization angle
\cite{milano2023_parkfrenet,milano2022_frenet_circuits}.
Geometric quantities have also been incorporated into control design
\cite{sanniti2022_curvature}, while the geometric interpretation of
instantaneous power establishes a multidimensional representation of the
non-active-power vector
\cite{milano2024_instpower}.

The three-phase voltage trajectory at the point of common coupling
(PCC) defines a tangent-normal-binormal (TNB) frame. In this frame, the converter
current components give distinct channels for active power and the
two transverse non-active-power components. This paper develops the
associated moving-frame model, decoupled current control, and analysis
of converter--grid interaction.

The main contributions of this paper are summarized as follows.
\begin{itemize}[leftmargin=*,topsep=2pt,itemsep=2pt,parsep=0pt,partopsep=0pt]

\item
\rev{A voltage-driven TNB current-control frame is constructed directly
from the instantaneous three-phase PCC-voltage trajectory.
The control axes require neither an explicit PLL synchronization angle
nor positive-, negative-, or zero-sequence decomposition.}

\item
\rev{Combined with the geometric instantaneous-power representation, the
TNB frame separates active-power transfer and the two transverse
non-active-power components into distinct orthogonal current channels,
providing a direct multidimensional power-regulation structure within
the same moving frame.}

\item
The converter $RL$-filter dynamics are derived in the moving TNB frame,
and curvature- and torsion-related coupling terms are explicitly
compensated to obtain decoupled nominal current dynamics.
\rev{A reduced transverse model is further developed to characterize
converter--grid interaction around a balanced operating point.}

\item
Controller-hardware-in-the-loop (CHIL) tests compare the proposed method
with PLL-based \rev{and PLL-less GFL} benchmarks
\rev{under matched plant, sampling, bandwidth, modulation, and
current-limiting conditions}.
The tests cover reference changes, voltage disturbances,
unbalance, \rev{and weak-grid operation}.

\end{itemize}

\begin{revision}
The remainder of this paper is organized as follows.
Section~\ref{sec:control_design} develops the voltage-driven Frenet frame,
TNB power representation, and moving-frame current controller.
Section~\ref{sec:ss_compare} develops the comparative dynamic and
grid-interaction analysis.
Section~\ref{sec:experimental_validation} presents the experimental
validation and benchmark comparisons.
Section~\ref{sec:conclusion} concludes the paper.
\end{revision}


\section{\rev{Voltage-Driven Frenet-Frame Current Control}}
\label{sec:control_design}

The controller uses a voltage-driven Frenet frame for GFL converters. Primes denote time derivatives,
$(\cdot)^\top$ denotes transpose, and $\norm{\cdot}$ is the Euclidean norm.

\subsection{Voltage-Driven Frenet Coordinates}
\label{subsec:frenet_coordinates}

Let $\bm v_p$ collect the instantaneous phase-to-neutral PCC voltages:
\begin{equation}
\bm v_p(t)
\equiv
\bm v_{p,abc}(t)
=
\begin{bmatrix}
v_{p,a}(t)&v_{p,b}(t)&v_{p,c}(t)
\end{bmatrix}^{\top}
\in\RR^3.
\label{eq:u_abc_def}
\end{equation}

\begin{revision}
Following the geometric-frequency formulation, associate
$\bm v_p$ with the integral curve
\begin{equation}
\bm x_p(t)
=
\bm x_p(t_0)
+
\int_{t_0}^{t}
\bm v_p(\sigma)\,\dd\sigma,
\qquad
\bm x_p'=\bm v_p.
\label{eq:voltage_integral_curve}
\end{equation}

Thus, the tangent of $\bm x_p$ is aligned with the instantaneous voltage,
rather than with the locus traced directly by $\bm v_p$.
\end{revision}

\rev{For twice-differentiable $\bm v_p$ satisfying
$\norm{\bm v_p}>0$ and
$\norm{\bm v_p\times\bm v_p'}>0$,
the tangent, binormal, and normal unit vectors are defined as}
\begin{equation}
\bm T
=
\frac{\bm v_p}{\norm{\bm v_p}},
\qquad
\bm B
=
\frac{
\bm v_p\times\bm v_p'
}{
\norm{\bm v_p\times\bm v_p'}
},
\qquad
\bm N
=
\bm B\times\bm T.
\label{eq:TNB_voltage}
\end{equation}

The triad $\{\bm T,\bm N,\bm B\}$ forms a right-handed orthonormal frame.
Its time-domain curvature- and torsion-related rates are
\begin{equation}
\omega_\kappa
=
\frac{
\norm{\bm v_p\times\bm v_p'}
}{
\norm{\bm v_p}^{2}
},
\qquad
\omega_\tau
=
\frac{
\bm v_p\cdot
\left(
\bm v_p'\times\bm v_p''
\right)
}{
\norm{\bm v_p\times\bm v_p'}^{2}
}
\norm{\bm v_p}.
\label{eq:omega_kappa_tau_voltage}
\end{equation}

Here, $\omega_\kappa$ governs the local rotation within the osculating
plane, whereas $\omega_\tau$ describes the evolution of that plane.

The Frenet transformation from $abc$ to TNB coordinates is
\begin{equation}
\Frenet
\triangleq
\begin{bmatrix}
\bm T^\top\\
\bm N^\top\\
\bm B^\top
\end{bmatrix},
\qquad
\Frenet^{-1}
=
\Frenet^\top.
\label{eq:F_def}
\end{equation}

For any vector $\bm y\in\RR^3$,
\begin{equation}
\bm y_{\tnb}
=
\Frenet\bm y,
\qquad
\bm y
=
\Frenet^\top\bm y_{\tnb}.
\label{eq:y_tnb_map}
\end{equation}

The Frenet--Serret kinematics give
\begin{equation}
\Frenet'
=
\OmegaF\Frenet,
\label{eq:F_time}
\end{equation}
where
\begin{equation}
\OmegaF
=
\begin{bmatrix}
0 & \omega_\kappa & 0\\
-\omega_\kappa & 0 & \omega_\tau\\
0 & -\omega_\tau & 0
\end{bmatrix}.
\label{eq:OmegaF_time}
\end{equation}

Since the tangential axis is aligned with the PCC voltage,
\begin{equation}
\bm v_{p,\mathrm{TNB}}
=
\Frenet\bm v_{p,abc}
=
\begin{bmatrix}
v_{p,\T}\\
0\\
0
\end{bmatrix},
\qquad
v_{p,\T}
=
\norm{\bm v_{p,abc}}.
\label{eq:8_u_tnb}
\end{equation}

The converter-current vector is
\begin{equation}
\bm i_{abc}
=
\begin{bmatrix}
i_a&i_b&i_c
\end{bmatrix}^{\top},
\end{equation}
and its TNB representation is
\begin{equation}
\bm i_{\mathrm{TNB}}
=
\Frenet\bm i_{abc}
=
\begin{bmatrix}
\iT\\
\iN\\
\iB
\end{bmatrix},
\label{eq:9_i_tnb}
\end{equation}
where $\iT$, $\iN$, and $\iB$ denote the tangential, normal, and binormal
current components, respectively.

The PCC voltage thus determines both the frame and its geometric rates
without an explicit synchronization angle.

\subsection{Instantaneous Power Decomposition}
\label{subsec:power_decomposition}

Following the instantaneous-power formulation in
\cite{milano2024_instpower}, the scalar active power and vector non-active power are defined as
\begin{equation}
P
=
\bm v_{p,abc}^{\top}\bm i_{abc},
\qquad
\bm Q
=
\bm v_{p,abc}\times\bm i_{abc}.
\label{eq:P_Q_abc}
\end{equation}

\rev{The vector $\bm Q$ is referred to as the non-active-power vector to
distinguish it from the scalar reactive power conventionally used under
balanced operation.}

Since $\Frenet$ is orthonormal, the inner product is invariant under the
coordinate transformation. Hence,
\begin{equation}
P
=
\bm v_{p,\mathrm{TNB}}^{\top}
\bm i_{\mathrm{TNB}}
=
v_{p,\T}\iT.
\label{eq:11_P_tnb}
\end{equation}

Using
\begin{equation}
\bm v_{p,abc}
=
v_{p,\T}\bm T,
\qquad
\bm i_{abc}
=
\iT\bm T
+
\iN\bm N
+
\iB\bm B,
\end{equation}
the non-active-power vector becomes
\begin{equation}
\bm Q
=
v_{p,\T}
\left(
\iN\bm B
-
\iB\bm N
\right).
\label{eq:12_Qvec_tnb}
\end{equation}

Accordingly, two orthogonal components of $\bm Q$ are
\begin{equation}
\begin{aligned}
Q_B
&\triangleq
\bm Q^\top\bm B
=
v_{p,\T}\iN,
\\
Q_{\N}
&\triangleq
-\bm Q^\top\bm N
=
v_{p,\T}\iB.
\end{aligned}
\label{eq:12_QB_QN}
\end{equation}

\begin{figure}[!t]
\centering
\plotbox{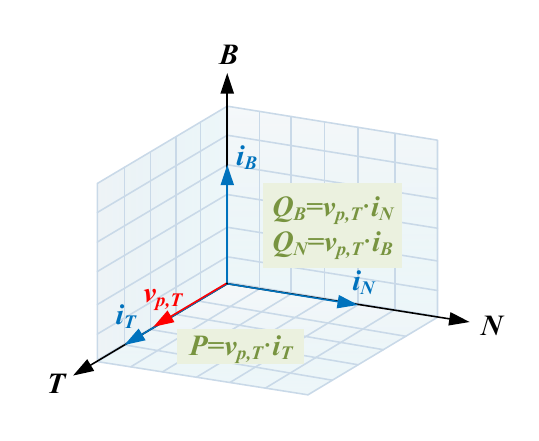}{0.70}
\caption{Geometric interpretation of instantaneous power in the
voltage-driven TNB coordinates.}
\label{fig:power_tnb}
\end{figure}

\begin{figure}[!t]
\centering
\plotbox{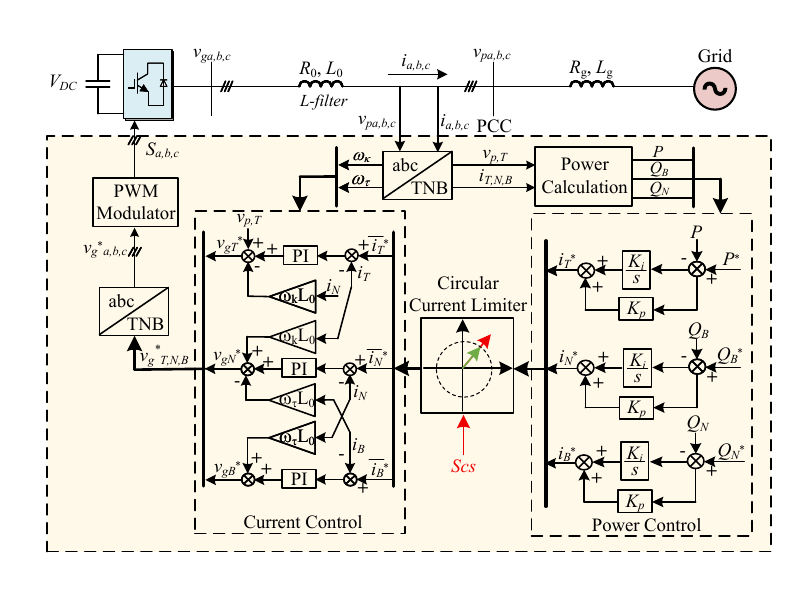}{1}
\caption{Block diagram of the proposed Frenet-frame current control scheme.}
\label{fig:control_block}
\vspace{-3mm}
\end{figure}

The signs in \eqref{eq:12_QB_QN} follow from the orientation of the
frame. The dot product retains the tangential current because the PCC
voltage is aligned with $\bm T$, while the cross product satisfies
$\bm T\times\bm N=\bm B$ and $\bm T\times\bm B=-\bm N$. With the stated
sign convention, $P$, $Q_B$, and $Q_N$ equal $v_{p,\T}$ times $\iT$,
$\iN$, and $\iB$, respectively, as illustrated in
Fig.~\ref{fig:power_tnb}. The control scheme in Fig.~\ref{fig:control_block}
uses this correspondence to assign each power reference to its
associated current channel. Proportional-integral (PI) regulators track
the current references, and the resulting converter-voltage commands
are implemented through pulsewidth modulation (PWM).

\subsection{Moving-Frame Model and Decoupled Current Control}
\label{subsec:moving_frame_control}

Each converter phase is connected to the PCC through an $RL$ output
filter with resistance $R_o$ and inductance $L_o$.
Let $\bm v_{g,abc}$ denote the converter-side phase voltages, expressed
with the same reference as $\bm v_{p,abc}$. For this symmetric per-phase
filter model, the voltage balance in $abc$ coordinates is
\begin{equation}
L_o\bm i_{abc}'
=
\bm v_{g,abc}
-
\bm v_{p,abc}
-
R_o\bm i_{abc}.
\label{eq:30_kvl_abc}
\end{equation}

From the product rule and \eqref{eq:F_time},
$\Frenet\bm i_{abc}'=\bm i_{\tnb}'-\OmegaF\bm i_{\tnb}$.
Premultiplying \eqref{eq:30_kvl_abc} by $\Frenet$ then gives
\begin{equation}
L_o\bm i_{\tnb}'
=
\bm v_{g,\tnb}
-
\bm v_{p,\tnb}
-
R_o\bm i_{\tnb}
+
L_o\OmegaF\bm i_{\tnb}.
\label{eq:kvl_tnb_matrix}
\end{equation}

Using
$\bm v_{p,\tnb}=[v_{p,\T}\ 0\ 0]^\top$,
the componentwise dynamics are
\begin{equation}
\left\{
\begin{aligned}
L_o\iT'
&=
v_{g,\T}
-
v_{p,\T}
-
R_o\iT
+
\omega_\kappa L_o\iN,
\\
L_o\iN'
&=
v_{g,\N}
-
R_o\iN
-
\omega_\kappa L_o\iT
+
\omega_\tau L_o\iB,
\\
L_o\iB'
&=
v_{g,\B}
-
R_o\iB
-
\omega_\tau L_o\iN.
\end{aligned}
\right.
\label{eq:14_kvl_tnb_comp}
\end{equation}

Let $v_{\PI,\T}$, $v_{\PI,\N}$, and $v_{\PI,\B}$ denote the outputs of the
PI current regulators driven by
$\iT^\star-\iT$, $\iN^\star-\iN$, and $\iB^\star-\iB$, respectively.
The converter-voltage references are designed as
\begin{equation}
\left\{
\begin{aligned}
v_{g,\T}^\star
&=
v_{p,\T}
+
v_{\PI,\T}
-
\omega_\kappa L_o\iN,
\\
v_{g,\N}^\star
&=
v_{\PI,\N}
+
\omega_\kappa L_o\iT
-
\omega_\tau L_o\iB,
\\
v_{g,\B}^\star
&=
v_{\PI,\B}
+
\omega_\tau L_o\iN.
\end{aligned}
\right.
\label{eq:vg_decouple}
\end{equation}

With exact feedforward and realizable voltage commands,
\eqref{eq:vg_decouple} and \eqref{eq:14_kvl_tnb_comp} give identical
nominal plants on the three axes:
\begin{equation}
L_o i_k'=v_{\PI,k}-R_o i_k,\qquad k\in\{\mathrm{T,N,B}\}.
\label{eq:decoupled_plants}
\end{equation}

The frame derivative in \eqref{eq:kvl_tnb_matrix} produces two coupling
pairs. Curvature couples the tangential and normal channels, while
torsion couples the normal and binormal channels. The feedforward
commands include the PCC voltage on the tangential axis and cancel
$L_o\OmegaF\bm i_{\tnb}$ across the three axes. The resulting identical
nominal plants allow the same current regulator to be used on each axis.

\subsection{\rev{Current- and Power-Loop Parameter Design}}
\label{subsec:controller_tuning}

Each decoupled current channel in \eqref{eq:decoupled_plants} has the plant
$1/(L_os+R_o)$. The PI regulator $G_i(s)=K_{p,i}+K_{i,i}/s$ cancels the
plant pole and sets the closed-loop bandwidth to $\omega_{ci}$:
\begin{equation}
K_{p,i}=L_o\omega_{ci},\qquad K_{i,i}=R_o\omega_{ci}.
\label{eq:current_pi_gains}
\end{equation}

The nominal closed-loop response is
$G_{ci}(s)=\omega_{ci}/(s+\omega_{ci})$. All three controllers use
$f_{ci}=\omega_{ci}/(2\pi)=1$~kHz in the comparative tests.

For $\bm p=[P\ Q_B\ Q_N]^\top$, linearization at the operating
point $(V_{T0},\bm i_{\mathrm{TNB},0})$ gives
\begin{equation}
\Delta\bm p=V_{T0}\Delta\bm i_{\mathrm{TNB}}
+\bm i_{\mathrm{TNB},0}\Delta v_{p,T}.
\label{eq:outer_power_linearization}
\end{equation}

The gain design holds $v_{p,T}$ at $V_{T0}$. The outer-loop crossover
is chosen to satisfy $\omega_o\leq0.1\omega_{ci}$, with
$f_o=\omega_o/(2\pi)=100$~Hz in the tests. For
$G_o(s)=K_{p,o}(1+\omega_{z,o}/s)$ with
$\omega_{z,o}=5$~rad/s, unity open-loop gain at $\omega_o$ yields
\begin{equation}
\begin{aligned}
K_{p,o}
&=\frac{1}{V_{T0}|G_{ci}(\mathrm{j}\omega_o)|
  \sqrt{1+(\omega_{z,o}/\omega_o)^2}},\\
K_{i,o}&=\omega_{z,o}K_{p,o}.
\end{aligned}
\label{eq:outer_kp}
\end{equation}

Uppercase $K$ denotes a continuous-domain gain, and lowercase $k$ in
Table~\ref{tab:sim_param_compare} denotes a normalized digital signal
processor (DSP) coefficient.

The second-order measurement filter attenuates switching ripple before
differentiation while retaining the voltage dynamics required for
reconstruction. The differentiator cutoff $f_f$ is kept above the
geometric-rate bandwidth $f_d$ to limit attenuation and phase lag in
the reconstruction band, and below one tenth of the sampling frequency
to limit high-frequency gain. A higher $f_d$ speeds up reconstruction
but passes more derivative noise to the feedforward terms. The settings
in Table~\ref{tab:sim_param_compare} place $f_d$ one decade below $f_f$
and $f_{ci}$.


\section{\rev{Comparative Dynamic and Grid-Interaction Analysis}}
\label{sec:ss_compare}

This section compares the decoupled phase-locked loop (D-PLL) and
Frenet controllers under matched plant and current-loop conditions. The reduced analysis examines in-plane
transverse coupling around a balanced operating point, synchronization
response, PLL-bandwidth sensitivity, and weak-grid interaction.

\subsection{D-PLL Benchmark Model}
\label{subsec:dpll_dynamics}

The positive- and negative-sequence decoupling stage of the D-PLL
benchmark is represented by the equivalent sequence-extraction filter
\begin{equation}
G_{\mathrm{seq}}(s)
=
\frac{\omega_{\mathrm{seq}}}
{s+\omega_{\mathrm{seq}}},
\label{eq:dpll_sequence_filter}
\end{equation}
where $\omega_{\mathrm{seq}}=2\pi f_{\mathrm{seq}}$. The PLL regulator is
\begin{equation}
G_{\mathrm{DPLL}}(s)
=
K_p^{\mathrm{DPLL}}
+
\frac{K_i^{\mathrm{DPLL}}}{s}.
\label{eq:dpll_pi}
\end{equation}

\begin{revision}
For a phase-detector gain $V_0$, the proportional and integral gains
set the second-order design targets before sequence-filter dynamics
are included ($G_{\mathrm{seq}}=1$):
\begin{equation}
K_p^{\mathrm{DPLL}}
=
\frac{
2\zeta_{\mathrm{PLL}}\omega_{n,\mathrm{PLL}}
}{V_0},
\qquad
K_i^{\mathrm{DPLL}}
=
\frac{
\omega_{n,\mathrm{PLL}}^2
}{V_0},
\label{eq:pll_gain_design}
\end{equation}
where $\omega_{n,\mathrm{PLL}}=2\pi f_{n,\mathrm{PLL}}$ and
$\zeta_{\mathrm{PLL}}$ are the design natural frequency and damping
ratio, respectively. In the PLL-bandwidth
sensitivity analysis, $\zeta_{\mathrm{PLL}}$ and the sequence-extraction
bandwidth are held constant, while the proportional and integral gains are
recalculated from \eqref{eq:pll_gain_design}.
\end{revision}

Around a balanced operating point, the transverse voltage perturbation
and estimated-angle perturbation satisfy
\begin{equation}
\begin{aligned}
\Delta e_{\mathrm{DPLL}}
&=
G_{\mathrm{seq}}
\left(
\Delta V_\perp
-
V_0\Delta\hat{\theta}
\right),
\\
\Delta\hat{\theta}
&=
\frac{G_{\mathrm{DPLL}}}{s}
\Delta e_{\mathrm{DPLL}},
\\
D_{\mathrm{DPLL}}(s)
&=
s+
V_0G_{\mathrm{DPLL}}G_{\mathrm{seq}}.
\end{aligned}
\label{eq:dpll_reduced_derivation}
\end{equation}

Eliminating the locking error and estimated-angle perturbation yields
\begin{equation}
\begin{aligned}
\Delta_{\mathrm{DPLL}}(s)
={}&
s^2(s+\omega_{\mathrm{seq}})
\\
&+
V_0\omega_{\mathrm{seq}}
\left(
K_p^{\mathrm{DPLL}}s
+
K_i^{\mathrm{DPLL}}
\right).
\end{aligned}
\label{eq:Delta_dpll}
\end{equation}

Hence, the sequence extractor and angle-locking loop introduce additional
synchronization modes into the D-PLL current-control path.

\subsection{Reduced Transverse Transfer Functions}
\label{subsec:reduced_models}

The two controllers use the same continuous-domain inner current regulator
\begin{equation}
G_i(s)=K_{p,i}+\frac{K_{i,i}}{s},
\qquad
D_i(s)=L_o s+R_o+G_i(s).
\label{eq:current_loop_compact}
\end{equation}

Let $I_0$ denote the operating current magnitude. For D-PLL, the
reduced transverse PCC-voltage-to-current response is
\begin{equation}
Y_{\mathrm{DPLL}}^{\perp}(s)
=
-
\frac{
G_{\mathrm{DPLL}}G_{\mathrm{seq}}
\left(
V_0+L_oI_0s
\right)
}{
D_iD_{\mathrm{DPLL}}
}.
\label{eq:Ydpll_final}
\end{equation}

Thus, the sequence-extraction and angle-locking dynamics enter directly
into the transverse current channel.

\begin{figure}[!t]
\centering
\plotbox{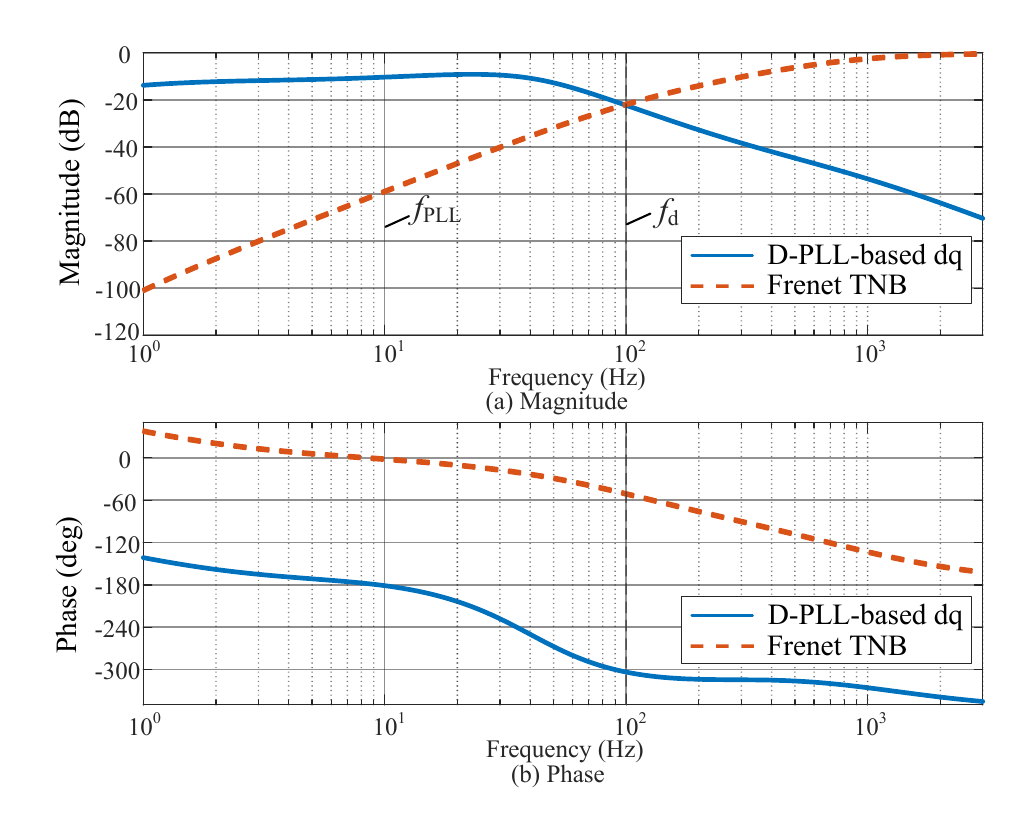}{0.98}
\caption{\rev{Normalized transverse output-admittance characteristics.
(a) Magnitude. (b) Phase.}}
\label{fig:freq_domain_compare}
\vspace{-2mm}
\end{figure}

\begin{revision}
For the Frenet controller, consider a balanced planar operating trajectory
with constant voltage magnitude $V_0$. A small angular perturbation
satisfies
\begin{equation}
\Delta V_\perp
=
V_0\Delta\theta,
\qquad
\Delta\omega_\kappa
=
s\Delta\theta.
\end{equation}

The corresponding ideal transverse voltage-to-geometric-rate relation is
therefore
\begin{equation}
K_{\kappa,\mathrm{id}}(s)
\triangleq
\frac{
\Delta\omega_\kappa(s)
}{
\Delta V_\perp(s)
}
=
\frac{s}{V_0}.
\label{eq:Kkappa_ideal}
\end{equation}

Let $D_f(s)=\omega_f s/(s+\omega_f)$ and
$G_r(s)=\omega_d/(s+\omega_d)$ denote the differentiator and
geometric-rate filter, with $\omega_f=2\pi f_f$ and
$\omega_d=2\pi f_d$. With residual delay $T_d$, the corresponding
transverse reconstruction model is
\begin{equation}
\widehat K_\kappa(s)
=
\frac{
G_r(s)D_f(s)e^{-sT_d}
}{
V_0
},
\label{eq:Kkappa_estimated}
\end{equation}
and the reconstruction residual becomes
\begin{equation}
E_\kappa(s)
=
K_{\kappa,\mathrm{id}}(s)
-
\widehat K_\kappa(s)
=
\frac{
s-G_r(s)D_f(s)e^{-sT_d}
}{
V_0
}.
\label{eq:Kkappa_implemented}
\end{equation}

The resulting estimator-inclusive transverse admittance is
\begin{equation}
Y_{\mathrm F}^{\perp}(s)
=
-
\frac{
L_oI_0E_\kappa(s)
}{
D_i(s)
}.
\label{eq:Yf_estimator_complete}
\end{equation}

The residual $E_\kappa(s)$ measures the difference between the ideal
and reconstructed geometric-rate responses to the same transverse
voltage perturbation. In \eqref{eq:Yf_estimator_complete},
$L_o I_0 E_\kappa(s)$ gives the corresponding coupling-voltage response,
and $D_i(s)$ relates that voltage to the transverse current.

Equations~\eqref{eq:Ydpll_final} and
\eqref{eq:Yf_estimator_complete} thus preserve the dominant
synchronization and reconstruction dynamics of the two control
architectures within the same reduced-order framework.

The responses in Fig.~\ref{fig:freq_domain_compare} are normalized by the
balanced operating-point admittance $I_0/V_0$, where $V_0$ and $I_0$
denote the Euclidean magnitudes of the balanced three-phase voltage and
current vectors, respectively. Under balanced operation,
$V_0=V_{LL,\mathrm{rms}}$ and
$I_0=\sqrt{3}I_{\mathrm{ph,rms}}$, where the subscript $\mathrm{rms}$
denotes root mean square (RMS). At 22~kW and zero non-active power,
$V_{LL,\mathrm{rms}}=380$~V gives
$V_0=380$~V and $I_0=57.89$~A.

The nominal D-PLL uses
$f_{n,\mathrm{PLL}}=20$~Hz,
$\zeta_{\mathrm{PLL}}=0.9$, and an
80-Hz sequence-extraction bandwidth. The Frenet implementation uses $f_d=100$~Hz, $f_f=1$~kHz, and one
sampling period of residual digital delay.
\end{revision}

\begin{revision}
As shown in Fig.~\ref{fig:freq_domain_compare}, the D-PLL controller
exhibits a pronounced low-frequency transverse-admittance peak associated
with the sequence-extraction and angle-locking dynamics. Over
1--100~Hz, its maximum normalized magnitude is approximately
$-9.12$~dB, compared with $-21.93$~dB for the Frenet controller,
corresponding to an approximately 12.8-dB reduction in the dominant
low-frequency transverse coupling.

The improvement is frequency dependent. The two magnitude characteristics
intersect near 99~Hz, close to the selected geometric-rate bandwidth.
Below this range, the Frenet reconstruction residual remains small,
whereas at higher frequencies the differentiator, geometric-rate filter,
digital delay, and current-loop dynamics increasingly determine the
response. The proposed controller therefore primarily suppresses the
low-frequency synchronization interaction rather than reducing the
transverse admittance over the entire frequency range.
\end{revision}

\subsection{Synchronization-Related Response Comparison}
\label{subsec:sync_response_compare}

The D-PLL angle-tracking channel is
\begin{equation}
C_{\mathrm{DPLL}}(s)
=
\frac{
V_0
G_{\mathrm{DPLL}}(s)
G_{\mathrm{seq}}(s)
}{
D_{\mathrm{DPLL}}(s)
}.
\label{eq:Cdpll_tracking}
\end{equation}

\begin{figure}[!t]
\centering
\plotbox{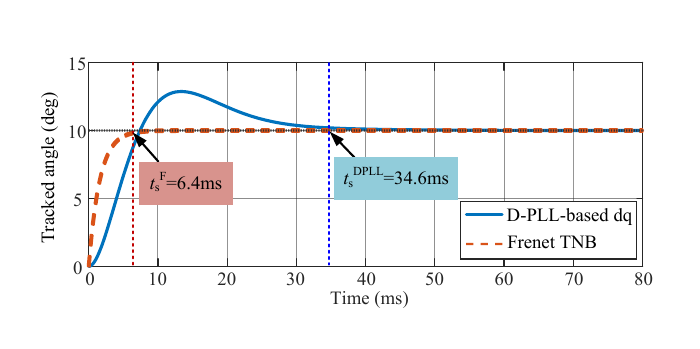}{1}
\caption{\rev{Equivalent synchronization responses to a
$10^\circ$ PCC-voltage angle step.}}
\label{fig:sync_response_compare}
\vspace{-2mm}
\end{figure}

\begin{revision}
For the Frenet controller, an equivalent angular-response channel can be
constructed from the reconstructed geometric rate as
\begin{equation}
C_{\mathrm F}(s)
=
\frac{
G_r(s)D_f(s)e^{-sT_d}
}{s}.
\label{eq:Cf_tracking}
\end{equation}

This channel provides an equivalent measure of the response time associated
with geometric-rate reconstruction and does not represent an internal
angle state of the Frenet controller.

Fig.~\ref{fig:sync_response_compare} compares the two responses to the
same $10^\circ$ PCC-voltage angle step.
\end{revision}

\begin{revision}
The nominal D-PLL response is underdamped and reaches the 2\% settling
band in approximately 34.6~ms. The equivalent Frenet response settles in
approximately 6.4~ms, consistent with the first-order settling-time
estimate $4/\omega_d=6.37$~ms at $f_d=100$~Hz. The latter response
therefore reflects the implemented estimator dynamics, including the
finite reconstruction bandwidth and digital delay.
\end{revision}

\subsection{\rev{PLL-Bandwidth Sensitivity}}
\label{subsec:pll_bandwidth_sweep}

\begin{revision}
The influence of the D-PLL bandwidth is assessed by varying its natural
frequency while maintaining
$\zeta_{\mathrm{PLL}}=0.9$,
$f_{\mathrm{seq}}=80$~Hz, and unchanged current-loop parameters.
Table~\ref{tab:pll_bandwidth_sweep} summarizes the maximum normalized
transverse-admittance magnitude over 1--100~Hz and the corresponding
2\% angle-tracking settling time.
\end{revision}

\begin{table}[!t]
\centering
\caption{\rev{Sensitivity to D-PLL Bandwidth}}
\label{tab:pll_bandwidth_sweep}
{\color{black}
\footnotesize
\renewcommand{\arraystretch}{1.16}
\setlength{\tabcolsep}{3.5pt}

\begin{tabular}{%
>{\centering\arraybackslash}p{0.18\columnwidth}
>{\centering\arraybackslash}p{0.45\columnwidth}
>{\centering\arraybackslash}p{0.26\columnwidth}}
\toprule
\shortstack{\textbf{$f_{n,\mathrm{PLL}}$}\\\textbf{(Hz)}} &
\shortstack{\textbf{Max. normalized}\\
\textbf{$|Y_{\mathrm{DPLL}}^\perp|$ (dB)}} &
\shortstack{\textbf{2\% settling}\\
\textbf{time (ms)}} \\
\midrule

10 & $-9.85$ & 77.9 \\
20 & $-9.12$ & 34.6 \\
30 & $-7.87$ & 18.2 \\
50 & $-4.67$ & 29.3 \\

\bottomrule
\end{tabular}
}
\end{table}

\begin{revision}
Increasing the PLL natural frequency from 10 to 30~Hz shortens the
angle-tracking settling time but simultaneously increases the transverse
admittance. Further increasing the natural frequency to 50~Hz no longer
improves the settling time because of the stronger interaction with the
fixed sequence-extraction dynamics. The nominal 20-Hz setting therefore
provides a practical compromise among angle-tracking speed, sequence
filtering, and transverse coupling.

The results further indicate that the synchronization response cannot be
characterized by the current-loop bandwidth alone. For the D-PLL
controller, it is jointly shaped by the PLL and sequence-extraction
dynamics, whereas for the Frenet controller it is governed by the
derivative and geometric-rate reconstruction paths. These
method-specific dynamics are therefore preserved in the subsequent CHIL
evaluation, with the remaining converter and implementation conditions
kept common.
\end{revision}

\subsection{\rev{Weak-Grid Interaction}}
\label{subsec:weak_grid_interaction}

\begin{revision}
The converter--grid interaction is assessed through
\begin{equation}
\det
\left[
\bm I+
\bm Z_g(s)\bm Y_o(s)
\right]
=
0,
\label{eq:minor_loop}
\end{equation}
where $\bm Y_o(s)$ and $\bm Z_g(s)$ denote the converter output-admittance
and grid-impedance matrices, respectively. For the transverse scalar
reduction used in the subsequent sweep of the short circuit ratio (SCR),
\eqref{eq:minor_loop} becomes
\begin{equation}
1+
Z_g^{\perp}(s)
Y^{\perp}(s)
=
0.
\label{eq:scalar_minor_loop}
\end{equation}

The stronger low-frequency transverse coupling of the D-PLL controller shown in Fig.~\ref{fig:freq_domain_compare} becomes increasingly influential as the grid impedance rises.  The Frenet controller removes the PLL angle-locking path but does not eliminate converter--grid interaction.  Below the geometric-rate bandwidth, the interaction is dominated by the reconstruction residual; at higher frequencies, the differentiator, geometric-rate filter, digital delay, and current-loop dynamics become increasingly important.

The controller formulations are independent of online knowledge of $R_g$ and $L_g$. Grid strength is therefore varied directly through the external grid impedance in the subsequent SCR sweep.

The reduced models characterize the local small-signal converter--grid interaction around a balanced operating point and assume a stiff dc link and the adopted $RL$ output filter.  Switching-frequency effects, current limiting, modulation saturation, protection actions, and trajectory-validity transitions fall outside the scope of these models.  Section~\ref{subsec:weak_grid_validation} compares the critical SCRs predicted from the closed-loop pole trajectories with those identified in the CHIL tests.
\end{revision}

\color{black}
\section{\rev{CHIL Validation}}
\label{sec:experimental_validation}

\subsection{\rev{Experimental Platform and Comparison Conditions}}
\label{subsec:rtds_setup}

\begin{figure}[!t]
\centering
\plotbox{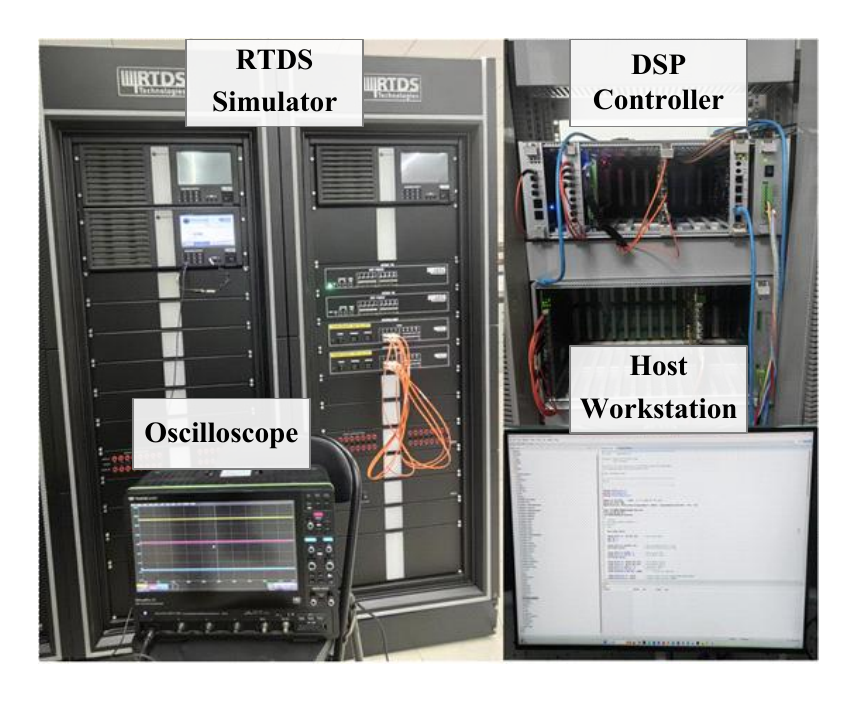}{0.90}
\caption{RTDS-based CHIL platform.}
\label{fig:rtds_platform}
\vspace{-2mm}
\end{figure}

The CHIL setup in Fig.~\ref{fig:rtds_platform} pairs a real-time digital
simulator (RTDS) with a TMS320C6657 DSP.
The case study uses a three-phase four-wire converter. The RTDS models
the converter, output filter, and Thevenin grid. It sends PCC voltages
and converter currents to the DSP, which
returns voltage commands at $f_s=15$~kHz, with $T_s=1/f_s$.
The grid impedance is $Z_g=R_g+\mathrm{j}\omega_0L_g$, where
$\omega_0=2\pi f_0$. Table~\ref{tab:sim_param_compare} lists the test
parameters.

\begin{revision}
The experimental comparison includes the D-PLL-based $dq$ controller,
LPV-PSC, and the proposed Frenet controller. All implementations use the
same converter, grid, operating point, sampling and switching frequencies,
measurement conditioning, modulation constraint, current limit, and
disturbance conditions.

LPV-PSC retains the published
gain-scheduled outer power-synchronization law
\cite{mansour2022_lpv_psc}, with the power-control bandwidth set to the
common 100-Hz target. The method-specific sequence-extraction,
gain-scheduled power-synchronization, and geometric-rate reconstruction
dynamics are retained in their respective implementations.
\end{revision}

\begin{table}[!t]
\caption{\rev{Plant and Controller Parameters}}
\label{tab:sim_param_compare}
\centering
\scriptsize
\renewcommand{\arraystretch}{1.025}
\setlength{\tabcolsep}{2.4pt}

\begin{tabularx}{\columnwidth}{@{}
>{\raggedright\arraybackslash}X
>{\centering\arraybackslash}p{0.23\columnwidth}
>{\centering\arraybackslash}p{0.29\columnwidth}
@{}}
\toprule
\textbf{Parameter} &
\textbf{Symbol} &
\textbf{Value}
\\
\midrule

\multicolumn{3}{@{}l}{\textit{Common plant and operating conditions}}
\\
\addlinespace[0.8mm]

Grid voltage and frequency &
$V_{LL},\,f_0$ &
$380$ V, $50$ Hz
\\

DC-link voltage &
$V_{dc}$ &
$1300$ V
\\

Switching \rev{and sampling} frequencies &
$f_{\rm sw},\,\rev{f_s}$ &
$15$ kHz, \rev{$15$ kHz}
\\

Output-filter parameters &
$R_o,\,L_o$ &
$0.02~\Omega$, $4$ mH
\\

Nominal grid parameters &
$R_g,\,L_g$ &
$0.0464~\Omega$, $6.960$ mH
\\

Active-power reference &
$P^\star$ &
$22$ kW
\\

\midrule
\multicolumn{3}{@{}l}{\textit{Matched current- and power-loop settings}}
\\
\addlinespace[0.8mm]

Outer-loop DSP PI coefficients &
$k_{p,P},\,k_{i,P}$ &
$0.1,\,0.01$
\\

Inner-loop DSP PI coefficients &
$k_{p,i},\,k_{i,i}$ &
$30,\,3$
\\

\rev{Current bandwidth and power crossover} &
\rev{$f_{ci},\,f_o$} &
\rev{$1$ kHz, $100$ Hz}
\\

\rev{Common rate-deviation bound} &
\rev{$|\Delta f|_{\max}$} &
\rev{$5$ Hz}
\\

\midrule
\multicolumn{3}{@{}l}{\textit{D-PLL controller}}
\\
\addlinespace[0.8mm]

Reactive-power reference &
$Q^\star$ &
$0$ kVAr
\\

D-PLL code-domain PI gains &
\rev{$k_{p,\rm PLL},\,k_{i,\rm PLL}$} &
$0.24,\,2.4$
\\

\rev{PLL natural frequency and sequence bandwidth} &
\rev{$f_{n,\rm PLL},\,f_{\rm seq}$} &
\rev{$20$ Hz, $80$ Hz}
\\

\rev{D-PLL damping ratio} &
\rev{$\zeta_{\rm PLL}$} &
\rev{$0.9$}
\\

\midrule
\multicolumn{3}{@{}l}{\rev{\textit{LPV-PSC controller}}}
\\
\addlinespace[0.8mm]

\rev{Reactive-power reference} &
\rev{$Q^\star$} &
\rev{$0$ kVAr}
\\

\rev{Power-control bandwidth} &
\rev{$f_{\rm PSC}$} &
\rev{$100$ Hz}
\\

\midrule
\multicolumn{3}{@{}l}{\textit{Frenet controller}}
\\
\addlinespace[0.8mm]

Non-active-power references &
$Q_B^\star,\,Q_N^\star$ &
$0,\,0$ kVAr
\\

\rev{Measurement-filter settings} &
\rev{$f_a,\,n_a$} &
\rev{$1.5$ kHz, $2$}
\\

\rev{Differentiator cutoff} &
\rev{$f_f$} &
\rev{$1$ kHz}
\\

\rev{Geometric-rate bandwidth} &
\rev{$f_d$} &
\rev{$100$ Hz}
\\

\bottomrule
\end{tabularx}
\end{table}

\begin{revision}
The current- and power-loop parameters follow
Section~\ref{subsec:controller_tuning}. The Frenet estimator settings
are listed in Table~\ref{tab:sim_param_compare}. The D-PLL gains are
obtained from \eqref{eq:pll_gain_design}. The common 5-Hz
rate-deviation bound is implemented as a protective saturation and does
not determine the nominal bandwidth of any controller.
\end{revision}

\subsection{\rev{Baseline Dynamic Comparison}}
\label{subsec:baseline_dynamic_tests}

\noindent\textit{1) Case 1: Power-Reference Tracking}

The first case evaluates active-power tracking at $\mathrm{SCR}=3$.
The reference is increased from $22$ to $25~\mathrm{kW}$ at
$t=1.0~\mathrm{s}$ and restored to $22~\mathrm{kW}$ at
$t=1.2~\mathrm{s}$, while the corresponding non-active-power references
remain zero. \rev{Active-power settling time is evaluated using a 2\% band
around the final steady-state mean.}

\begin{revision}
The non-active-power metric is $|Q|$ for D-PLL and LPV-PSC and
$\sqrt{Q_B^2+Q_N^2}$ for Frenet. The synchronization-related deviation
is reported in per unit (p.u.) as $(f-f_0)/f_0$ for the frequency
estimated by D-PLL and LPV-PSC and as
$(\omega_\kappa-\omega_0)/\omega_0$ for Frenet.
\end{revision}

\begin{figure}[!t]
\centering
\plotbox{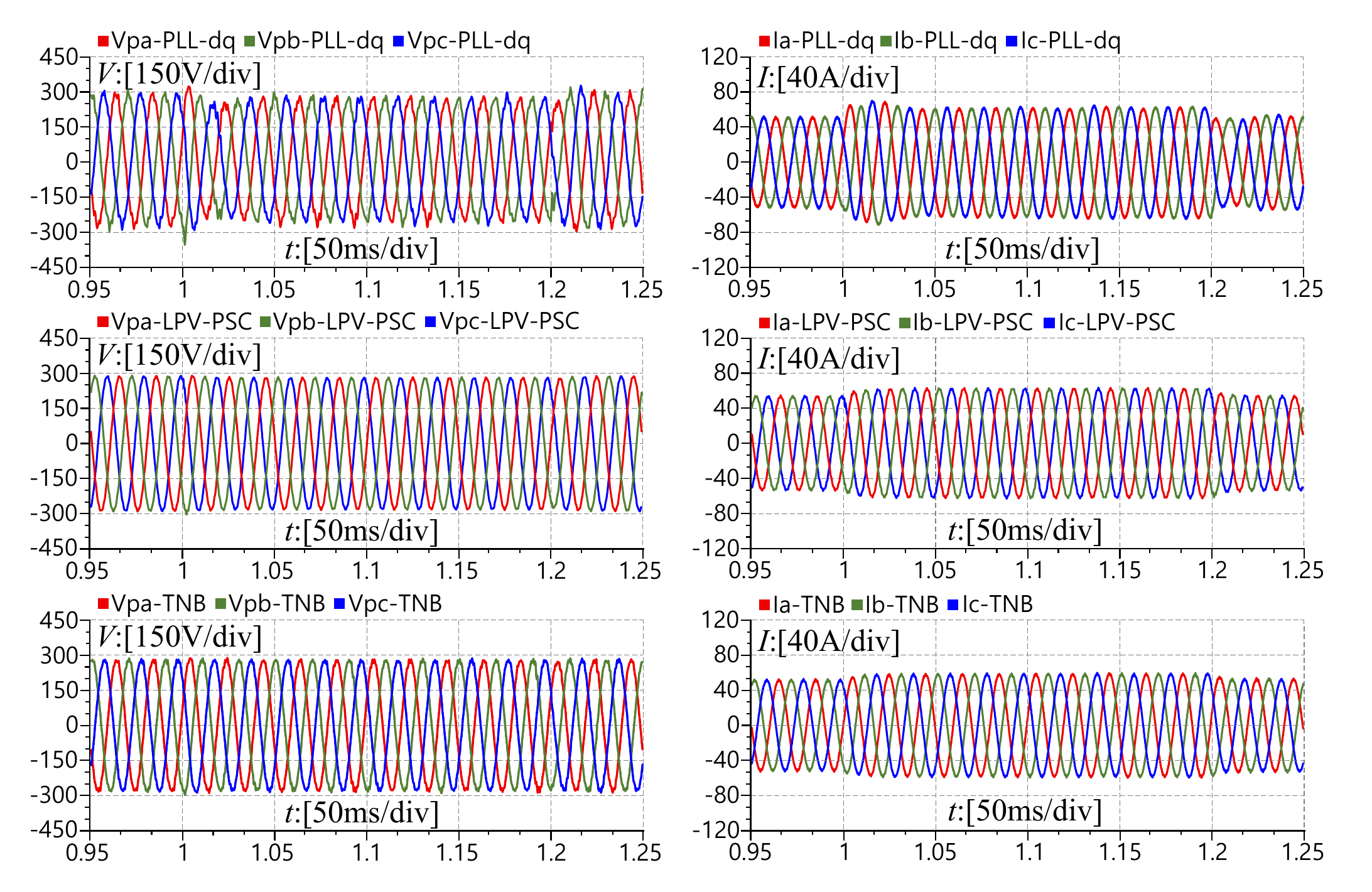}{1.00}
\caption{\rev{Three-phase PCC-voltage and converter-current responses
in Case~1.}}
\label{fig:case1_abc_waveforms}
\vspace{-2mm}
\end{figure}

\begin{figure}[!t]
\centering
\plotbox{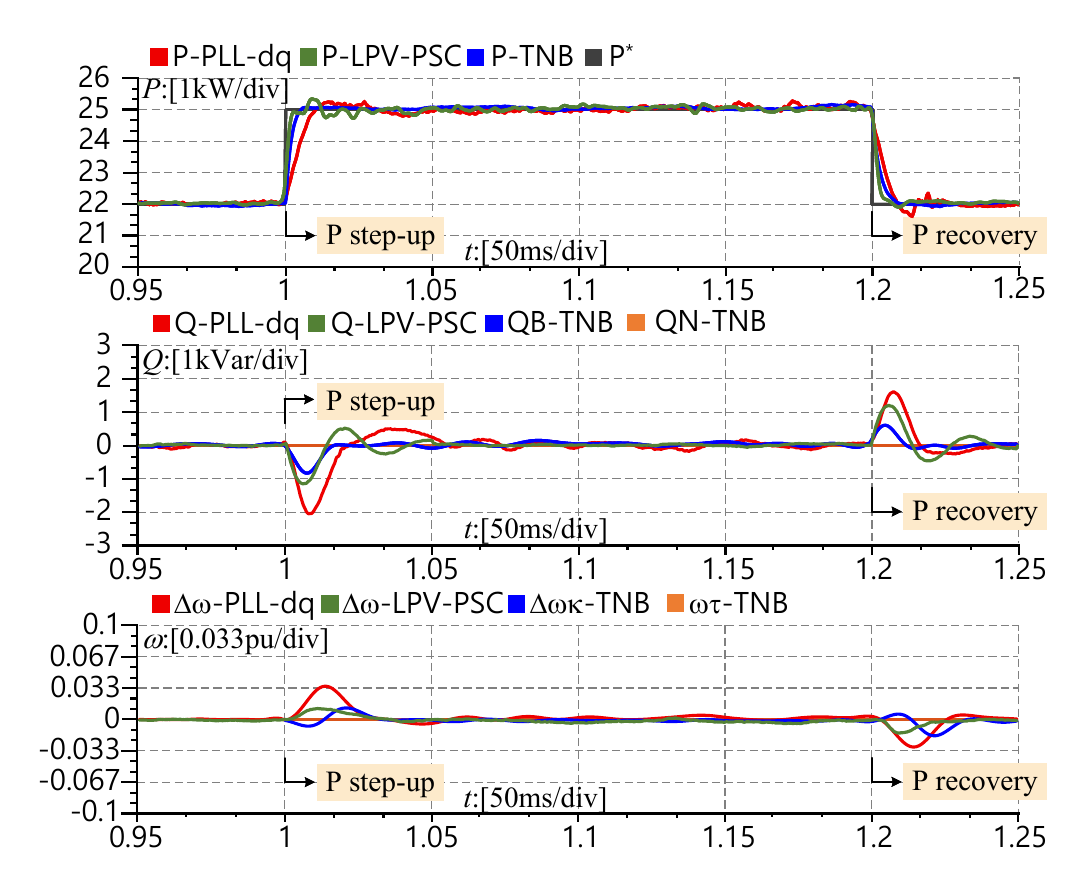}{0.98}
\caption{\rev{Active-power, non-active-power, and synchronization-related
rate responses in Case~1.}}
\label{fig:case1_pqf_waveforms}
\vspace{-2mm}
\end{figure}

\begin{revision}
Fig.~\ref{fig:case1_abc_waveforms} shows that the phase-current amplitudes
increase after the upward power step and return near their initial levels
when the reference is restored. For Frenet, \eqref{eq:11_P_tnb} relates
this change in power to the tangential current.

As shown in Fig.~\ref{fig:case1_pqf_waveforms}, all three controllers
track both reference changes. D-PLL rises more slowly and briefly
undershoots at recovery. The signed non-active-power excursions reverse
direction between transitions, with $Q_B$ dominant for Frenet.
For the upward step, Table~\ref{tab:rtds_perf_summary} reports settling
times of 20, 10, and 6~ms for D-PLL, LPV-PSC, and Frenet, respectively.
Frenet gives the smallest non-active-power excursion, 0.84~kVAr,
compared with 1.12~kVAr for LPV-PSC and 2.05~kVAr for D-PLL. The peak
normalized rate deviation is 0.011~p.u. for both PLL-less controllers
and 0.033~p.u. for D-PLL.
\end{revision}

\noindent\textit{2) Case 2: Balanced Grid Disturbance}
\label{subsec:case_balanced_disturbance}

A balanced $20^\circ$ voltage-angle step is applied at
$t=1.0~\mathrm{s}$. The source voltage magnitude is then reduced from
1.0 to 0.8~p.u. at $t=1.3~\mathrm{s}$ and restored at $t=1.6~\mathrm{s}$. The voltage-driven trajectory remains planar, so this case primarily distinguishes the synchronization and reconstruction dynamics.

\begin{figure}[!t]
\centering
\plotbox{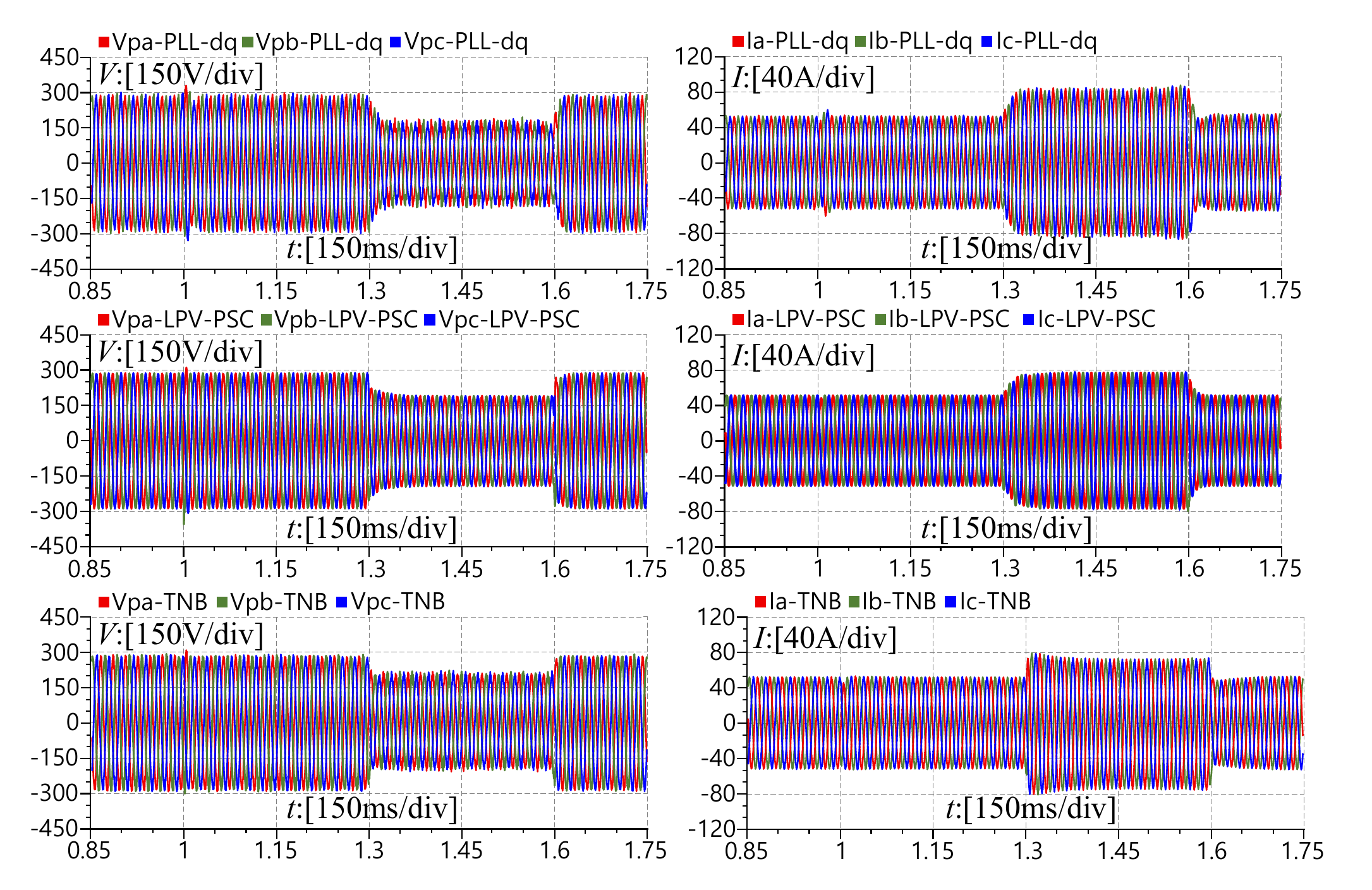}{0.98}
\caption{\rev{Three-phase PCC-voltage and converter-current responses
in Case~2.}}
\label{fig:case2_vi_waveforms}
\vspace{-2mm}
\end{figure}

\begin{figure}[!t]
\centering
\plotbox{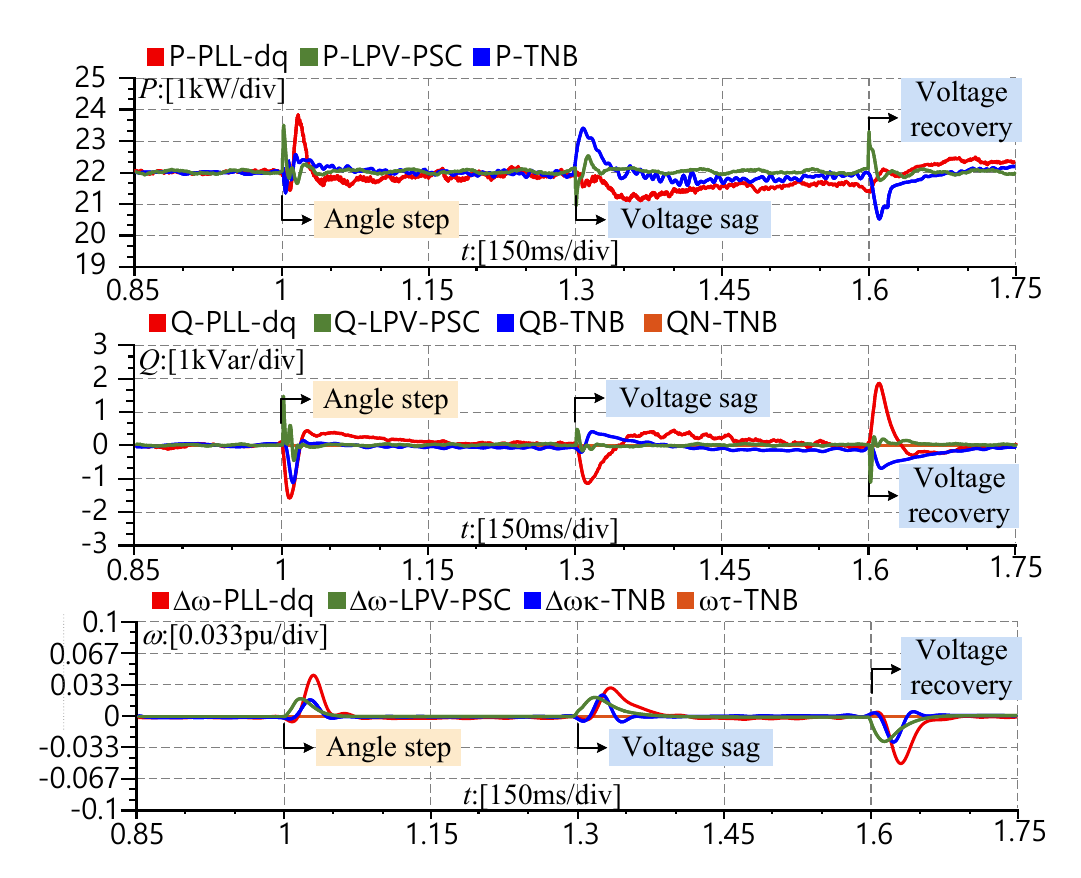}{0.98}
\caption{\rev{Active-power, non-active-power, and synchronization-related
rate responses in Case~2.}}
\label{fig:case2_pqf_waveforms}
\vspace{-2mm}
\end{figure}

\begin{revision}
Fig.~\ref{fig:case2_vi_waveforms} shows that the current amplitudes
return near their pre-disturbance levels after the angle step. During
the sag, the phase-current amplitudes increase for all three controllers
as the PCC voltage falls. The active power returns near its pre-sag level while the
current remains elevated, consistent with the higher tangential
current required by \eqref{eq:11_P_tnb} at lower voltage. The current
amplitudes decrease as the voltage recovers.

Fig.~\ref{fig:case2_pqf_waveforms} shows a pronounced D-PLL active-power
overshoot after the angle step and its largest non-active-power and
rate excursions at voltage recovery. Frenet active power rises briefly at
sag onset and dips at voltage recovery. The torsion-related rate stays
close to zero, consistent with the planar trajectory.
Over the full sequence, Table~\ref{tab:rtds_perf_summary} reports the
smallest peak active- and non-active-power deviations for Frenet, at
1.45~kW and 1.10~kVAr, respectively. Both PLL-less controllers reach
0.026~p.u. peak normalized rate deviation, compared with 0.050~p.u.
for D-PLL.
\end{revision}

\noindent\textit{3) Case 3: Single-Phase Voltage Unbalance}
\label{subsec:case_single_phase_unbalance}

\begin{revision}
A 0.5-p.u. magnitude reduction is applied to phase A of the Thevenin-source
voltage from $t=1.0$ to $1.4~\mathrm{s}$, reducing its magnitude from
1.0 to 0.5~p.u., while the other two phase magnitudes remain unchanged.
The resulting single-frequency unbalanced voltage trajectory remains
approximately planar but becomes noncircular, producing sustained
oscillatory power and TNB-current components.
\end{revision}

\begin{figure}[!t]
\centering
\plotbox{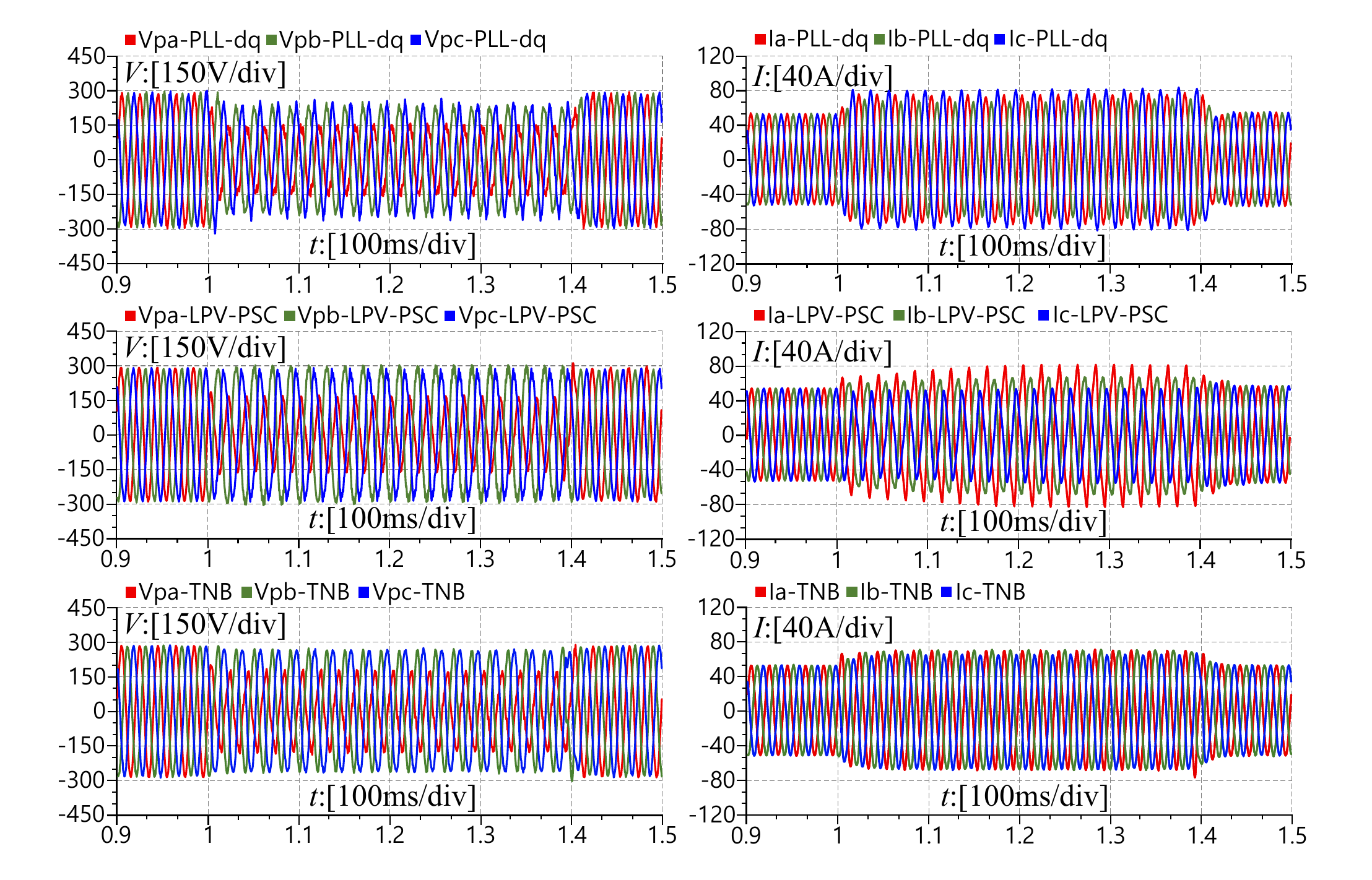}{0.98}
\caption{\rev{Three-phase PCC-voltage and converter-current responses
under the single-phase voltage unbalance in Case~3.}}
\label{fig:case3_abc_waveforms}
\vspace{-2mm}
\end{figure}

\begin{figure}[!t]
\centering
\plotbox{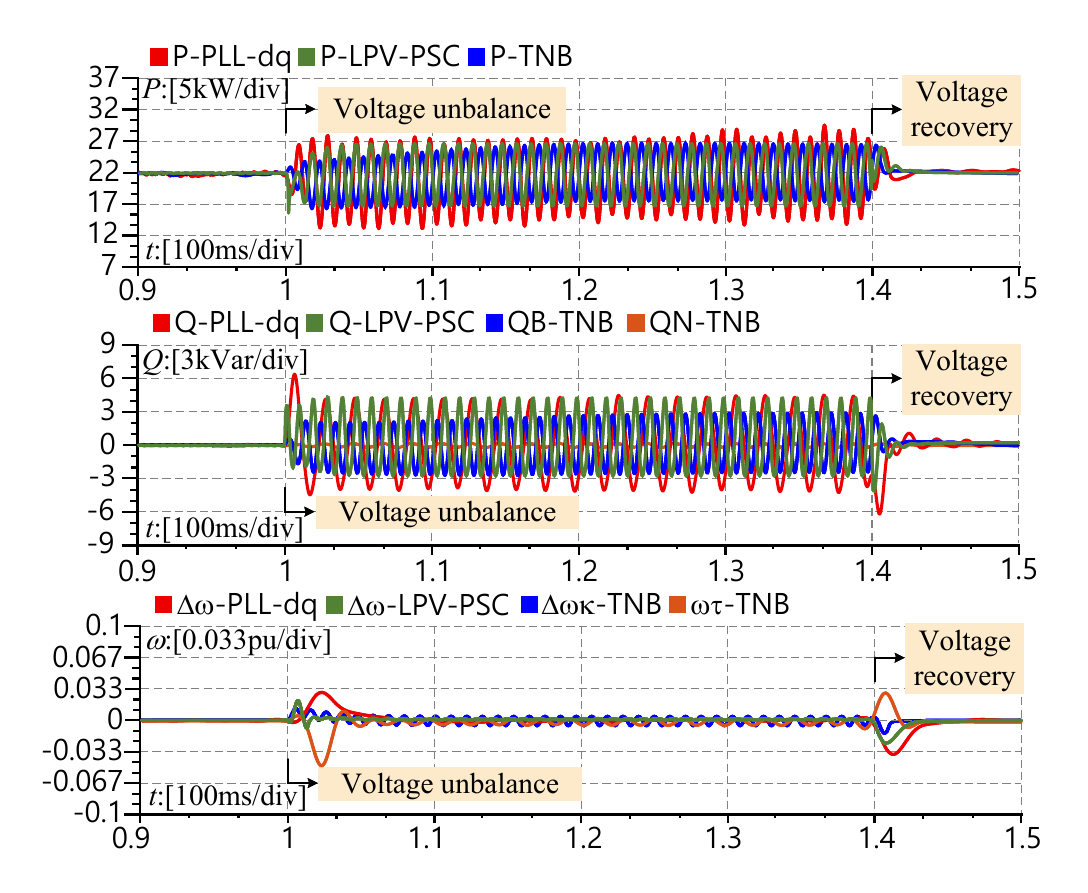}{0.92}
\caption{\rev{Active-power, non-active-power, and synchronization-related
rate responses in Case~3.}}
\label{fig:case3_pqf_waveforms}
\vspace{-2mm}
\end{figure}

\begin{figure}[!t]
\centering
\plotbox{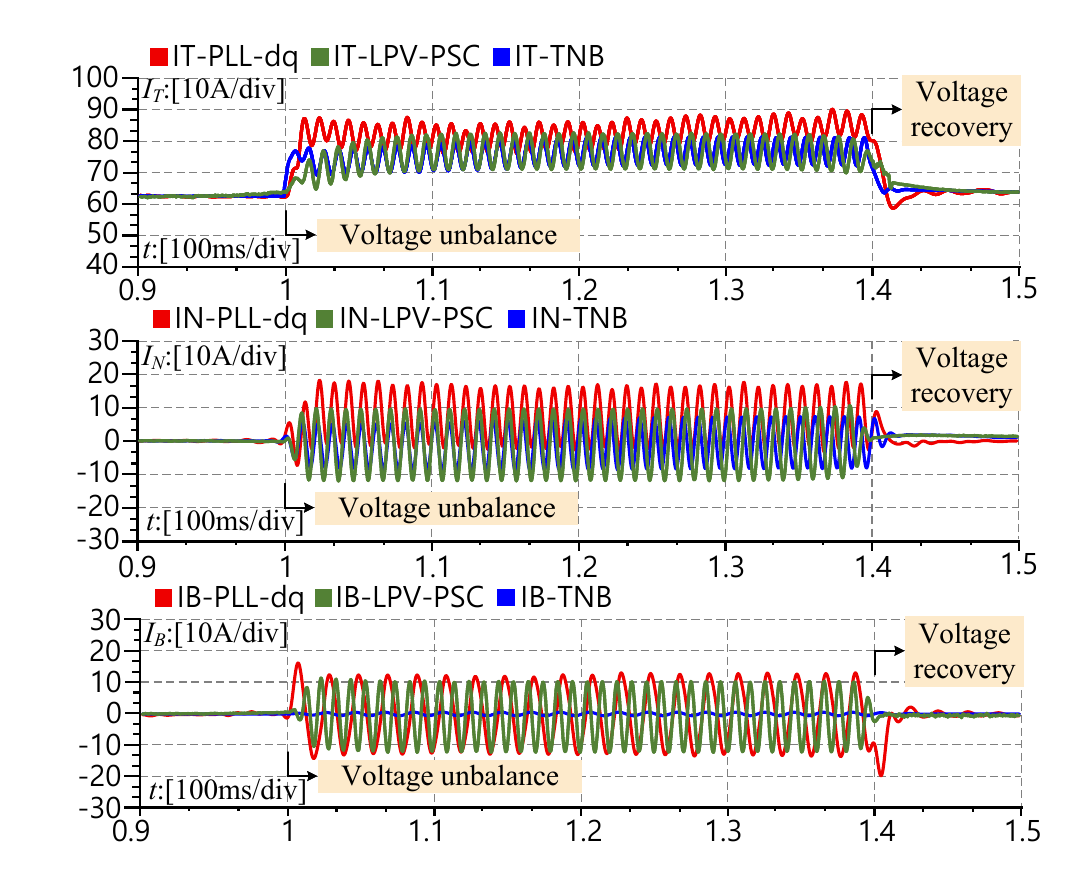}{0.92}
\caption{\rev{Currents of the three controllers resolved in the common
voltage-driven TNB frame in Case~3.}}
\label{fig:case3_tnb_current_components}
\vspace{-2mm}
\end{figure}

\begin{revision}
As shown in Fig.~\ref{fig:case3_abc_waveforms}, all three controllers
maintain bounded phase currents during the disturbance.
Fig.~\ref{fig:case3_pqf_waveforms} shows the largest non-active-power
excursions for D-PLL. For Frenet, $Q_B$ dominates the non-active-power
response, with a plotted peak $|Q_N|$ of about 0.22~kVAr. The
normalized torsion-related rate $\omega_\tau/\omega_0$ stays close to
zero during the sag and varies
mainly at its onset and clearance.

Fig.~\ref{fig:case3_tnb_current_components} resolves all phase currents in
the same voltage-driven TNB frame. Frenet gives the smallest peak
$|i_B|$, with a plotted value of 0.75~A.
Table~\ref{tab:rtds_perf_summary} reports Case~3 peaks over 1.0--1.4~s
and RMS values over 1.15--1.35~s from the plotted traces. Over these
intervals, the peak and RMS non-active-power magnitudes are 2.9 and
1.9~kVAr for Frenet, 6.4 and 2.9~kVAr for D-PLL, and 4.3 and 2.5~kVAr
for LPV-PSC.
\end{revision}

\begin{figure}[!t]
\centering
\plotbox{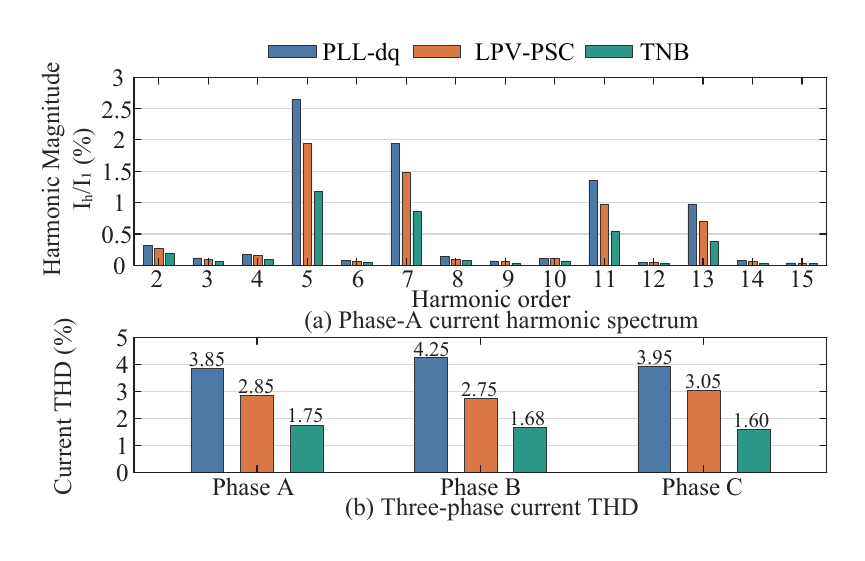}{0.96}
\caption{\rev{Current-quality comparison in Case~3.
(a) Phase-A harmonic spectrum from the second to fifteenth orders.
(b) THD of each phase current using all resolved harmonics.}}
\label{fig:case3_current_spectrum}
\vspace{-2mm}
\end{figure}

\begin{revision}
Current total harmonic distortion (THD) over 1.15--1.35~s is $100\sqrt{\sum_{h=2}^{H} I_h^2}/I_1$, where $I_h$ is the RMS current of harmonic $h$ and $H$ is the highest resolved order. Panel (a) of Fig.~\ref{fig:case3_current_spectrum} displays the second through fifteenth harmonics.
\end{revision}

\begin{revision}
The fifth and seventh harmonics dominate the displayed spectrum.  The maximum phase-current THD is $4.25\%$ for D-PLL, $3.05\%$ for LPV-PSC, and $1.75\%$ for Frenet.
\end{revision}

\begin{table*}[!t]
\centering
\caption{\rev{Representative CHIL Performance Comparison}}
\label{tab:rtds_perf_summary}

\scriptsize
\renewcommand{\arraystretch}{1.025}
\setlength{\tabcolsep}{4.0pt}

\renewcommand{\tabularxcolumn}[1]{m{#1}}

\begin{tabularx}{\textwidth}{
>{\centering\arraybackslash}m{0.065\textwidth}
>{\centering\arraybackslash}m{0.145\textwidth}
>{\centering\arraybackslash}X
>{\centering\arraybackslash}m{0.115\textwidth}
>{\centering\arraybackslash}m{0.115\textwidth}
>{\centering\arraybackslash}m{0.115\textwidth}
}

\toprule
\textbf{Case} &
\textbf{Test condition} &
\textbf{Performance index} &
\textbf{D-PLL} &
\rev{\textbf{LPV-PSC}} &
\textbf{Frenet}
\\
\midrule

\multirow{3}{*}{Case~1}
&
\multirow{3}{*}{\shortstack{
Upward active-power\\
reference step}}
&
Active-power settling time
&
$20$ ms
&
\rev{$10$ ms}
&
$6$ ms
\\

&
&
Peak non-active-power deviation
&
$2.05$ kVAr
&
\rev{$1.12$ kVAr}
&
$0.84$ kVAr
\\

&
&
\shortstack{Peak normalized synchronization-related\\rate deviation}
&
$0.033$~p.u.
&
\rev{$0.011$~p.u.}
&
$0.011$~p.u.
\\

\midrule

\multirow{3}{*}{Case~2}
&
\multirow{3}{*}{\shortstack{
Balanced phase-angle\\
step and voltage sag}}
&
Peak active-power deviation
&
$1.85$ kW
&
\rev{$1.51$ kW}
&
$1.45$ kW
\\

&
&
Peak non-active-power deviation
&
$1.89$ kVAr
&
\rev{$1.45$ kVAr}
&
$1.10$ kVAr
\\

&
&
\shortstack{Peak normalized synchronization-related\\rate deviation}
&
$0.050$~p.u.
&
\rev{$0.026$~p.u.}
&
$0.026$~p.u.
\\

\midrule

\multirow{4}{*}{Case~3}
&
\multirow{4}{*}{\shortstack{
Single-phase\\
voltage-unbalance\\
disturbance}}
&
Peak common-frame $|i_B|$
&
$16.1$ A
&
$12.4$ A
&
$0.75$ A
\\
&
&
Peak non-active-power magnitude
&
$6.4$ kVAr
&
$4.3$ kVAr
&
$2.9$ kVAr
\\
&
&
RMS non-active-power magnitude
&
$2.9$ kVAr
&
$2.5$ kVAr
&
$1.9$ kVAr
\\
&
&
\rev{Maximum phase-current THD}
&
\rev{$4.25\%$}
&
\rev{$3.05\%$}
&
\rev{$1.75\%$}
\\

\bottomrule
\end{tabularx}
\end{table*}

\subsection{\rev{Weak-Grid Stability Boundary}}
\label{subsec:weak_grid_validation}

\begin{revision}
Grid strength is characterized with
\begin{equation}
\mathrm{SCR}
=
\frac{V_{LL,b}^{2}}
{S_b
\sqrt{
R_g^2+
(\omega_0L_g)^2
}},
\label{eq:scr_definition}
\end{equation}
where $V_{LL,b}$ and $S_b$ denote the line-to-line voltage and apparent-power bases, respectively.  The nominal $X/R$ ratio is retained while $R_g$ and $L_g$ are scaled proportionally to obtain $\mathrm{SCR}\in\{5,\,3,\,2,\,1.5,\,1.2\}$.

At each grid strength, an independent CHIL grid-connection test is performed for each controller to determine whether it can establish the $P^\star=22~\mathrm{kW}$ steady operating point. No controller is redesigned or retuned for an individual SCR test point. A test point is classified as stable when grid connection is established, the phase currents remain bounded, and the active power enters and remains within the 2\% band around 22~kW without sustained post-connection oscillation. 
\end{revision}

\begin{figure}[!t]
\centering
\plotbox{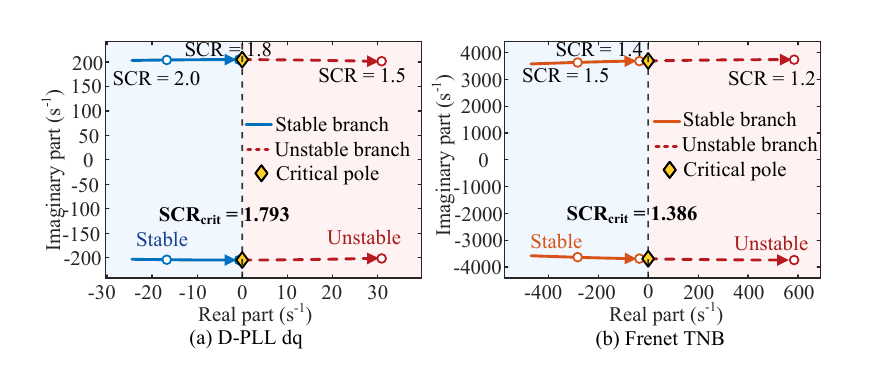}{1}
\caption{\rev{Closed-loop pole trajectories from the reduced weak-grid
models under decreasing SCR.
(a) D-PLL-based control.
(b) Frenet control.}}
\label{fig:theoretical_scr_boundary}
\end{figure}

\begin{revision}
Fig.~\ref{fig:theoretical_scr_boundary} gives critical SCR values of
1.793 and 1.386 for the D-PLL and Frenet controllers, respectively.
The corresponding critical oscillation frequencies are approximately
32.6 and 587.7~Hz. The lower critical SCR of the Frenet controller
indicates an extended weak-grid stability range. The limiting mode shifts
from the low-frequency PLL-dominated interaction toward a higher-frequency
range increasingly governed by the estimator, digital delay, and
current-loop dynamics.
\end{revision}

\begin{figure}[!t]
\centering
\plotbox{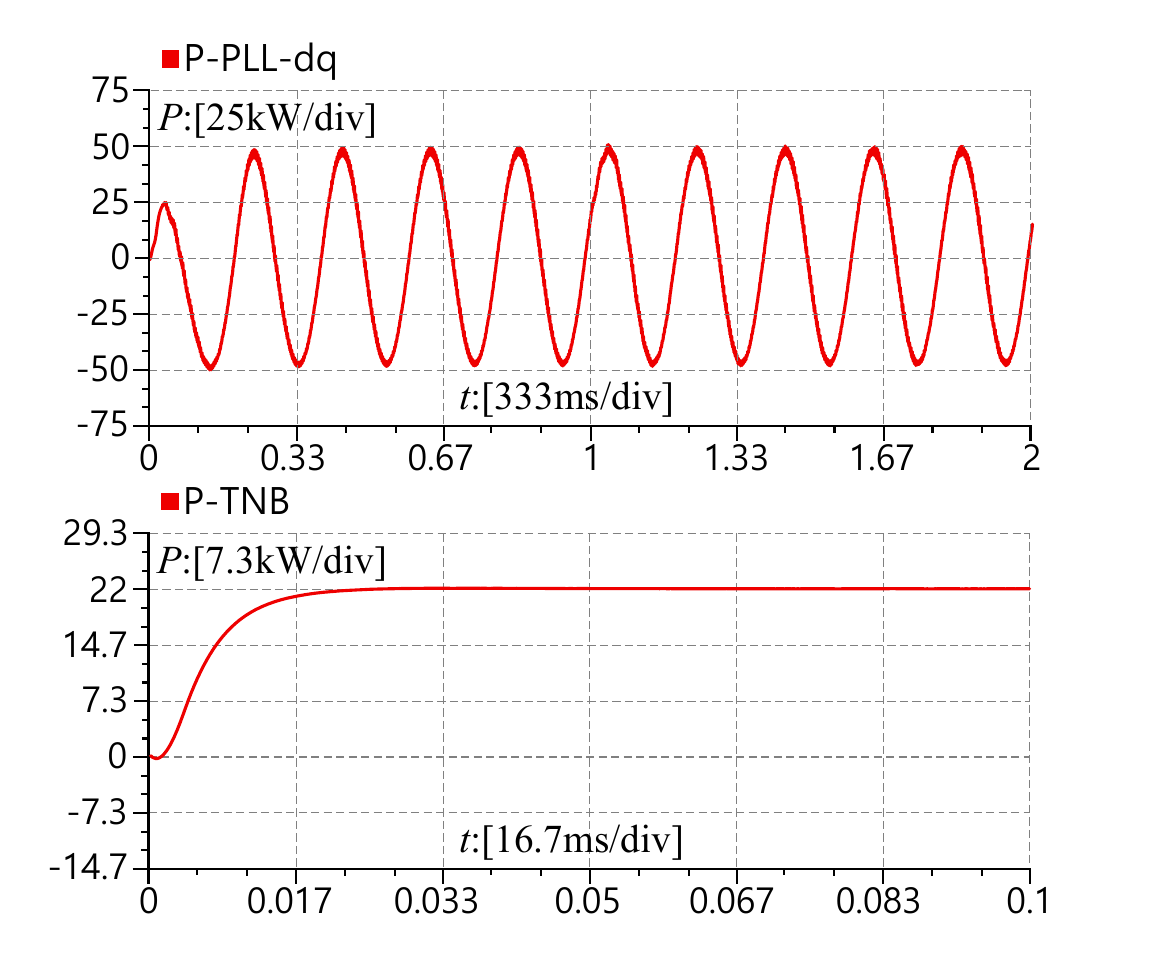}{0.85}
\caption{\rev{CHIL grid-connection active-power responses at
$\mathrm{SCR}=1.7$ for the D-PLL-based and Frenet controllers.}}
\label{fig:scr17_power_response}
\vspace{-2mm}
\end{figure}

\begin{table*}[!t]
\centering
\caption{\rev{Grid-Connection Stability and Active-Power Settling Time
Under Different SCRs}}
\label{tab:scr_results}

{\color{black}
\scriptsize
\renewcommand{\arraystretch}{1.025}
\setlength{\tabcolsep}{6.0pt}

\begin{tabularx}{\textwidth}{
>{\centering\arraybackslash}X
>{\centering\arraybackslash}p{0.085\textwidth}
>{\centering\arraybackslash}p{0.085\textwidth}
>{\centering\arraybackslash}p{0.085\textwidth}
>{\centering\arraybackslash}p{0.085\textwidth}
>{\centering\arraybackslash}p{0.085\textwidth}
>{\centering\arraybackslash}p{0.16\textwidth}}

\toprule

\multirow{2}{*}{\textbf{Controller}} &
\multicolumn{5}{c}{
\textbf{Grid-connection settling time at each SCR (ms)}
} &
\multirow{2}{*}{\textbf{Critical SCR}}
\\

\cmidrule(lr){2-6}

&
\textbf{5} &
\textbf{3} &
\textbf{2} &
\textbf{1.5} &
\textbf{1.2} &
\\

\midrule

D-PLL $dq$ &
$15$ &
$20$ &
$32$ &
U &
U &
$1.8$
\\

LPV-PSC &
$10$ &
$13$ &
$19$ &
$27$ &
U &
$1.4$
\\

Frenet &
$8$ &
$10$ &
$14$ &
$23$ &
U &
$1.4$
\\

\bottomrule
\end{tabularx}

\vspace{0.5mm}
\begin{minipage}{\textwidth}
\footnotesize
\raggedright
U denotes instability; the critical SCR is determined from a finer sweep.
\end{minipage}

}
\end{table*}

\begin{revision}
Table~\ref{tab:scr_results} gives CHIL critical SCRs of about 1.8 for
D-PLL and 1.4 for Frenet, close to the analytical values of 1.793 and
1.386, respectively. LPV-PSC exhibits a comparable experimental
stability boundary of approximately 1.4. Over the stable test points,
Frenet also retains shorter grid-connection settling times as the grid
weakens.

The agreement among the pole trajectories, the response at
$\mathrm{SCR}=1.7$ in Fig.~\ref{fig:scr17_power_response}, and the
CHIL sweep indicates that the reduced transverse model captures the
dominant local weak-grid interaction around the tested operating point. At
$\mathrm{SCR}=1.2$, all three controllers are unstable; lower SCR values
are therefore omitted from the principal sweep.
\end{revision}


\section{Conclusion}
\label{sec:conclusion}

This paper develops a PLL-less synchronization and current-control
method for GFL converters operating under weak and unbalanced grid
conditions. The method constructs a TNB frame directly from the
instantaneous PCC voltage. Aligning the tangent with the voltage
assigns active power to the tangential current and the two non-active-power
components to the normal and binormal currents.

The moving-frame filter model identifies the cross-axis voltage terms
introduced by curvature and torsion. Feedforward compensation yields
identical nominal plants on all axes, allowing the same inner-loop
tuning. A reduced transverse model relates the residual from
finite-bandwidth geometric-rate reconstruction to the current response.
For a $10^\circ$ voltage-angle step, the modeled equivalent
synchronization response settles in 6.4~ms with Frenet and 34.6~ms
with D-PLL. Over 1--100~Hz, Frenet reduces the peak normalized transverse
admittance by 12.8~dB relative to D-PLL.

In CHIL tests, Frenet settles in 6~ms after the active-power reference step,
compared with 10~ms for LPV-PSC and 20~ms for D-PLL. Frenet gives the
smallest non-active-power excursion in this test and the smallest peak
deviations in active and non-active power during the balanced angle step,
voltage sag, and recovery test. During single-phase voltage unbalance,
Frenet yields the lowest peak binormal current, 0.75~A, and the lowest
maximum phase-current THD, $1.75\%$. Its RMS non-active-power magnitude
is 1.9~kVAr, compared with 2.5~kVAr for LPV-PSC and 2.9~kVAr for D-PLL.

The model predicts critical SCRs of 1.386 for Frenet and 1.793 for
D-PLL. The CHIL sweep finds comparable critical SCRs for Frenet and
LPV-PSC, with shorter grid-connection settling times for Frenet at their
common stable test points. At $\mathrm{SCR}=1.7$, Frenet establishes the
22-kW operating point while the D-PLL power response continues to
oscillate. The pole-derived stability boundaries apply locally around
the balanced operating point used for linearization.

\FloatBarrier
\bibliographystyle{IEEEtran}
\bibliography{references}

\end{document}